# Regularity in the intratube interaction in double-wall carbon nanotubes

J. Arvanitidis[1], D. Christofilos[1,*], K. Papagelis[2], T. Takenobu[3], Y. Iwasa[3], H. Kataura[4], S. Ves[5], and G. A. Kourouklis[1]

[1]Physics Division, School of Technology, Aristotle University of Thessaloniki, 541 24 Thessaloniki, Greece

[2]Materials Science Department, University of Patras, 265 04 Patras, Greece

[3]Institute for Materials Research, Tohoku University, 2-1-1 Katahira, Aoba-ku, Sendai 980-8577 and CREST, Japan Science and Technology Corporation, Kawaguchi 332-0012, Japan

[4]National Institute of Advanced Industrial Science and Technology (AIST), 1-1-1 Higashi, Tsukuba, Ibaraki 305-8562, Japan

[5]Physics Department, Aristotle University of Thessaloniki, 541 24 Thessaloniki, Greece

High pressure Raman experiments were performed on bundled double wall carbon nanotubes (DWCNTs). We observe peculiar regularity in the strength of the inner-outer tube (intratube) interaction, which is attributed to the interplay between the tube size and the intratube spacing. Moreover, a pressure-induced red shift of the band gap energies of the probed inner tubes is deduced from the Raman spectra. These experimental findings reveal that the interaction between primary and secondary shells, although generally assumed to be weak, determines the dynamics in the interior of the DWCNTs.

Double-wall carbon nanotubes (DWCNTs), lying between single- (SWCNTs) and multi-wall carbon nanotubes (MWCNTs), constitute the most basic system for studying encapsulation effects in these systems. The Raman spectrum of carbon nanotubes allows selective probing of tubes for which the excitation wavelength is in resonance with the energy spacing between the corresponding van Hove singularities (sensitively dependent on the tube diameter and chirality) appearing on either side of the Fermi energy in the 1D electronic density of states (e-DOS).[1,2] Moreover, the frequencies of the low-energy Raman peaks, attributed to the radial breathing modes (RBMs) of the nanotubes, are inversely proportional to the tube diameter,[3] allowing the precise study of individual carbon nanotubes of different structural and electronic properties.

The unusually narrow RBMs appearing in the Raman spectrum of the DWCNTs reflect the growth of inner or secondary tubes with a high degree of perfection, being well protected in the interior of the outer or primary rolled graphene shells.[4] In this system, the existence of inner-outer tube (intratube) interaction causes the splitting of the Raman peaks, attributed to DWCNTs having the same secondary tube, but included in different diameter primary tubes.[5,6] In addition, the resulting frequencies for the RBMs of both inner and outer tubes are upshifted with decreasing interlayer distance (increasing intratube interaction).[5,7]

The application of high pressure on carbon nanotubes is a valuable tool for the investigation of their mechanical and structural stability.[8] In bundled SWCNTs, the van der Waals tube-tube (intertube) interaction dominates their pressure response, resulting in the nanotube cross-section deformation and eventually its collapse.[9,10,11] More specifically, a high pressure Raman study of a wide diameter range HiPCO (high pressure carbon monoxide)-derived SWCNT sample revealed that this deformation becomes more pronounced as the tube diameter increases.[12] In the case of DWCNT bundles, the intertube interaction is less effective as the secondary tubes enhance the structural rigidity of the outer shells, which, nevertheless, suffer considerable deformation at elevated pressure. On the other hand, the internal tubes are quasi-isolated due to their encapsulation inside the primary tubes.[13,14] This effect is clearly manifested by the normalized pressure derivatives of the RBM frequencies of the internal tubes, which take values close to that predicted theoretically for an isolated tube.[12,15] All these experimental findings provide a clear evidence for the reduction of the effective pressure exerted on the inner tubes through their relatively weak coupling with the outer ones. However, there is no clear picture concerning the strength of these intratube interactions as well as its relation to the characteristics of the specific tubes constituting a DWCNT. In this Letter, we report the results of a high pressure Raman study on a high quality bundled DWCNT material, which reveal novel aspects of the intratube interactions on the pressure response of the internal tubes. Surprisingly, smaller diameter secondary tubes can be more sensitive to pressure application depending on the encapsulating tube properties.

The DWCNT sample used in this study was prepared by the peapod conversion route, following Bandow's procedure[16]. Details of the sample preparation and the extensive characterization of the starting SWCNT material, the intermediate step peapods and the resulting DWCNTs by means of transmission electron microscopy (TEM), X-ray diffraction (XRD) and Raman spectroscopy can be found elsewhere.[13,17,18,19,20] The Raman spectra were recorded using a DILOR XY micro-Raman system equipped with a cryogenic charge coupled device (CCD) detector. High pressure was applied by means of a Mao-Bell type diamond anvil cell. The 4:1 methanol-ethanol mixture was used as pressure transmitting medium and the ruby fluorescence technique was used for pressure calibration. For excitation, the 514.5 nm line of an $Ar^+$ or the 647.1 and the 676.4 nm lines of a $Kr^+$ laser was focused on the sample by means of a 20x objective, while the laser power was kept below 2.5 mW - measured directly before the cell- in order to eliminate laser-heating effects.

Figure 1 shows the Raman spectra of the DWCNT material in the low frequency region, where the RBMs of the two concentric rolled graphene sheets constituting the DWCNTs are located, at room temperature and various pressures using the 676.4 nm line of a Kr$^+$ laser. The lower energy, broad band is attributed to the larger diameter primary tubes, while the narrow peaks at higher frequencies are assigned to the secondary tubes. The significantly large number of the observed RBM peaks, providing a wealth of information on the encaged tubes and their interaction, have been previously assigned to specific nanotubes by means of detailed Raman spectroscopic studies at low temperature and ambient pressure.[4,5,21] All RBM peaks of the present DWCNT material are blue-shifted with increasing pressure, reflecting the C-C bonds stiffening. In addition, lineshape as well as relative peak intensity changes occur at elevated pressures. Namely, the low energy band associated with radial vibrations of the primary tubes broadens and its intensity attenuates remarkably. These changes have been attributed to the pressure induced deformation of the outer tube cross section resulting from the intertube interaction within a bundle, earlier observed for both SWCNTs[9,10] and DWCNTs.[13] However, in the DWCNT material the primary tubes RBM band persists to much higher pressure than that in SWCNTs, suggesting -in agreement with our previous findings[13]- their enhanced structural stability provided by the secondary tubes.

The peaks attributed to the secondary tubes remain relatively narrow up to the highest pressure attained, reflecting that these tubes are essentially isolated from external perturbations and ruling out their pressure induced faceting analogous to that encountered for the primary tubes in a bundled sample. However, the Raman intensity distribution among these spectral features is gradually modified with pressure and two additional peaks appear at higher pressures in the frequency region 340-370 cm$^{-1}$. The internal tube RBM relative intensity profile in the Raman spectrum of the DWCNT material at 9.4 GPa excited by the 676.4 nm ($E_{exc}$= 1.833 eV) line, resembles closely to the corresponding pattern recorded at ambient pressure with 647.1 nm ($E_{exc}$= 1.916 eV) excitation (top panel in figure 1). Therefore, the RBM intensity redistribution is due to the pressure-induced red shift of the band gap energies, changing the resonance conditions of the probed inner carbon nanotubes. In fact, there is experimental evidence that in individual SWCNTs the energy spacings between the van Hove singularities decrease upon hydrostatic pressure with rates lying in the range 0.3-4.6 meV/kbar.[22] As a result, at higher pressures, the experimental Raman relative intensities at a specific excitation energy will be favorable for those tubes that at ambient conditions have higher energy van Hove singularities and thus their Raman spectrum will be more intense if it is excited by a smaller wavelength. Indeed, this is exactly what is found in our study, implying that, although the internal are quasi-isolated by the external tubes, the intratube coupling plays a crucial role in their pressure response and thus cannot be neglected. On the other hand, the pressure induced shift in the resonance conditions of the external tubes alone cannot account for the observed strong intensity attenuation and the significant broadening of their RBM band at higher pressures. This is an unambiguous experimental evidence for the pressure induced cross-section deformation of the primary tubes in bundled DWCNTs.

Upon pressure release, the frequency shifts of the observed Raman peaks are fully reversible and the original peak positions in the spectrum are restored at 1 bar (bottom spectrum in figure 1). However, the direct comparison of the spectra at ambient conditions before and after the pressure cycling reveals that, although the relative intensities of the Raman peaks attributed to the secondary tubes are almost identical, the intensity of the primary tubes RBM band is attenuated and the peak remains broader after pressure release. These divergences agree well with the proposed residual pressure-induced deformations of the primary tubes in DWCNTs[13] and in analogy to those observed in bundled SWCNTs.[9]

A complete picture of the inter- and the intratube interactions within a bundled DWCNT material could be depicted by compiling and comparing the mode Gruneisen parameters, $\gamma_i$ {= $B(\partial(\ln\omega_i)/\partial P)$, where $B$ is the bulk modulus} of the peaks appearing in the Raman spectrum. In figure 2, the normalized pressure slopes, $\Gamma_i = (1/\omega_i)(\partial\omega_i/\partial P)$, proportional to the mode Gruneisen parameters, of the RBMs in the DWCNTs are plotted as a function of the peak position. The presented data are obtained with excitation at 514.5 nm and 676.4 nm, along with $\Gamma_i$ parameters taken from the literature for a wide diameter range HiPCO-derived SWCNT sample (excitation with 514.5 and 632.8 nm)[12] and a DWCNT material prepared also by the peapod conversion route (excitation with 514.5 and 647.1 nm).[15] The top axis diameter values are indicative, as they are estimated on the basis of the simple relation $\omega_{RBM}(cm^{-1}) = 234/d_t(nm) + 10$, where $\omega_{RBM}$ is the RBM frequency and $d_t$ the tube diameter. This relation has been applied previously for SWCNT[23] and DWCNT[4,13] bundles. However, this is only a rough estimation, as it does not take into account the intratube interactions, causing the splitting of the observed RBMs and their frequency upshift.[5,7] The experimental data of the present work, concerning the normalized pressure slopes of various primary and secondary tubes constituting the DWCNT material, are in excellent agreement with those previously reported.[12,15] The primary tubes of the DWCNTs, exhibit strong and similar frequency (and thus diameter) dependence to the bundled SWCNTs. Namely, the normalized pressure derivatives decrease quasi-quadratically as the tube diameter decreases, which is rationalized in terms of the enhanced cross section fragility of the larger nanotubes upon pressure.[12] The comparison of the experimentally obtained $\Gamma_i$ values for the SWCNTs and the external tubes of the DWCNTs with the theoretically predicted value of the $\Gamma$ parameter for an isolated SWCNT (dotted horizontal line in figure 2),[12] reveal that the intertube van der Waals interactions govern the pressure response of these tubes. On the contrary, all data related to the secondary tubes lie close to that expected for an isolated tube with single exception being the half-filled green circle for which an unambiguous assignment cannot be made.[13] The observed quasi-isolation of the secondary tubes is in line with the previously reported data.[15] However, in the present work almost all the possible internal tubes in the diameter range 0.6-0.9 nm are observed as well as several split components, allowing the elucidation of the systematics governing the intratube interactions.

The $\Gamma_i$ values associated with the internal tubes are shown in the inset of figure 2. Although, these data lie around the value predicted for an isolated tube, there is an obvious overall trend of reducing $\Gamma$ with increasing frequency (decreasing nanotube diameter). This suggests that the primary tubes' size-dependent cross section deformation is "transferred" through the van der Waals intratube interactions to the secondary ones. Under this consideration, extending the pressure range of the experiments may finally cause a deformation of the internal tubes, similar to that suffered by the externals.

The most remarkable point arising from the RBM pressure derivatives presented here is their clustering in quasi-linear distributions in which the above mentioned $\Gamma_i$-$d_t$ relation is inverted. Note that $\Gamma_i$ values found by Venkateswaran[15] lie on the extrapolation of the groups illustrated in figure 2. In order to understand this observation, we invoke some simple geometric considerations regarding the relative size of the tubes constituting a DWCNT. The chiral vectors (n,m) of the carbon nanotubes quantizes their possible diameter values, leading to discrete intratube spacings. Depending on the chiral vectors involved, inner-outer tube combinations can be found where the minimum interlayer spacing approaches the "best fit" condition (close to the turbostratic constraint of graphite), whereas other combinations deviating from this condition, result in DWCNTs with higher intratube distance.[7] This is manifested by the intratube spacing dispersion of 0.34-0.38 nm observed in TEM experiments.[4] Moreover, for a specific internal tube, several compatible externals may exist, leading to very close intratube spacings, but enough to cause an observable split of the RBMs in the Raman spectrum. In the split components, the better the tube matching, the higher the frequency and the higher the corresponding $\Gamma_i$. As an example, the split components located at ambient pressure in the frequency region 361-367 cm$^{-1}$, which are attributed to the (8,1) nanotube,[4] follow exactly this scheme with pressure. Nevertheless, according to Pfeiffer et al.,[5] the experimentally observed RBM frequency splitting due to secondary tube enclosure in different primary tubes does not exceed 5 cm$^{-1}$. Therefore, the extended range of the grouping presented in figure 2 (exceeding 30 cm$^{-1}$) can be only partially interpreted in this way. Consider now that in figure 2 a point adjacent to a group of split components corresponds to a smaller internal tube (higher Raman frequency). If this tube has a smaller intratube distance, then the corresponding $\Gamma_i$ value will be larger than that of the lower frequency neighboring group of split components (the data belong in the same quasi-linear group in figure 2). Alternatively, an increase of the inner-outer tube mismatch will lead to smaller $\Gamma_i$ value (the data belong in different quasi-linear groups in figure 2).

Taking into account all the above considerations, each distribution of the RBM pressure derivatives can be understood as grouping of data related i) to the same secondary tube included into different diameter primary tubes, as well as ii) appropriate inner-outer tube combinations in such a way that their mutual interaction increases quasi-linearly with the reduction of the secondary tube diameter. For a quantitative interpretation of the observed ordering in the $\Gamma_i$ parameters, detailed theoretical investigations are required. Nevertheless, our experimental study strongly suggests that the pressure response of the internal tubes in DWCNTs is crucially dependent on the intratube interaction, which, in turn, is determined by the chiral vectors and the resulting intratube spacing of the inner-outer shell combination.

This work was partly supported by NEDO and MEXT, Japan. The financial support by a M. Curie reintegration grant (MERG-CT-2004-513606) to J. A. is also acknowledged.

# References

*Author to whom correspondence should be addressed; electronic mail: christof@vergina.eng.auth.gr

[1] M. S. Dresselhaus and P. C. Eklund, Adv. Phys. **49**, 705 (2000).

[2] A. Jorio, A.G. Souza Filho, G. Dresselhaus, M. S. Dresselhaus, A. K. Swan, M. S. Unlu, B. B. Goldberg, M. A. Pimenta, J. H. Hafner, C. M. Lieber, and R. Saito, Phys. Rev. B **65**, 155412 (2002).

[3] A. M. Rao, E. Richter, S. Bandow, B. Chase, P. C. Eklund, K. A. Williams, S. Fang, K. R. Subbaswamy, M. Menon, A. Thess, R. E. Smalley, G. Dresselhaus, and M. S. Dresselhaus, Science **275**, 187 (1997).

[4] R. Pfeiffer, H. Kuzmany, Ch. Kramberger, Ch. Schaman, T. Pichler, H. Kataura, Y. Achiba, J. Kurti, and V. Zolyomi, Phys. Rev. Lett. **90**, 225501 (2003).

[5] R. Pfeiffer, Ch. Kramberger, F. Simon, H. Kuzmany, V. N. Popov, and H. Kataura, Eur. Phys. J. B **42**, 345 (2004).

[6] M. Xia, S. Zhang, X. Zuo, E. Zhang, S. Zhao, L. Zhang, Y. Liu, and R. Liang, Phys. Rev. B **70**, 205428 (2004).

[7] A. Rahmani, J.-L. Sauvajol, J. Cambedouzou, and C. Benoit, Phys. Rev. B **71**, 125402 (2005).

[8] I. Loa, J. Raman Spectrosc. **34**, 611 (2003).

[9] U. D. Venkateswaran, A. M. Rao, E. Richter, M. Menon, A. Rinzler, R. E. Smalley, and P. C. Eklund, Phys. Rev. B **59**, 10928 (1999).

[10] M. J. Peters, L. E. McNeil, J. P. Lu, D. Kahn, Phys. Rev. B **61**, 5939 (2000).

[11] J. Tang, L. C. Qin, T. Sasaki, M. Yudasaka, A. Matsushita, and S. Iijima, Phys. Rev. Lett. **85**, 1887 (2000).

[12] U. D. Venkateswaran, D. L. Masica, G. U. Sumanasekera, C. A. Furtado, U. J. Kim, and P. C. Eklund, Phys. Rev. B **68**, 241406 (2003).

[13] J. Arvanitidis, D. Christofilos, K. Papagelis, K. S. Andrikopoulos, T. Takenobu, Y. Iwasa, H. Kataura, S. Ves, and G. A. Kourouklis, Phys. Rev. B **71**, 125404 (2005).

[14] P. Puech, H. Hubel, D. J. Dunstan, R. R. Bacsa, C. Laurent, and W. S. Bacsa, Phys. Rev. Lett. **93**, 095506 (2004).

[15] U. D. Venkateswaran, Phys. Stat. Sol. (b) **241**, 3345 (2004).

[16] S. Bandow, M. Takizawa, K. Hirahara, M. Yudasaka, and S. Iijima, Chem. Phys. Lett. **337**, 48 (2001).

[17] H. Kataura, Y. Kumazawa, Y. Ohtsuka, S. Suzuki, Y. Maniwa, and Y. Achiba, Synth. Metals **103**, 2555 (1999).

[18] H. Kataura, Y. Maniwa, T. Kodama, K. Kikuchi, H. Hirahara, K. Suenaga, S. Iijima, S. Suzuki, Y. Achiba, and W. Kratschmer, Synth. Metals **121**, 1195 (2001).

[19] H. Kataura, Y. Maniwa, M. Abe, A. Fujiwara, T. Kodama, K. Kikuchi, H. Imahori, Y. Misaki, S. Suzuki, and Y. Achiba, Appl. Phys. A: Mater. Sci. Process. **74**, 349 (2002).

[20] M. Abe, H. Kataura, H. Kira, T. Kodama, S. Suzuki, Y. Achiba, K. Kato, M. Takata, A. Fujiwara, K. Matsuda, and Y. Maniwa, Phys. Rev. B **68**, 041405 (2003).

[21] Ch. Kramberger, R. Pfeiffer, H. Kuzmany, V. Zolyomi, and J. Kurti, Phys. Rev. B **68**, 235404 (2003).

[22] J. Wu, W. Walukiewicz, W. Shan, E. Bourret-Courchesne, J. W. Ager III, K. M. Yu, E. E. Haller, K. Kissell, S. M. Bachilo, R. B. Weisman, and R. E. Smalley, Phys. Rev. Lett. **93**, 017404 (2004).

[23] A. Jorio, M. A. Pimenta, A. G. Souza Filho, R. Saito, G. Dresselhaus, and M. S. Dresselhaus, New J. Phys. **5**, 139 (2003).

**Figure Captions**

**Figure 1.** Raman spectra of the DWCNTs in the RBM frequency region at room temperature and for various pressures, recorded upon pressure increase and after total pressure release (bottom spectrum). The 676.4 nm line of a $Kr^+$ laser was used for excitation. Arrows mark Raman peaks appearing in the spectrum at elevated pressure. For comparison, the Raman spectrum of the DWCNT material at ambient conditions and using the 647.1 nm line of the $Kr^+$ laser is also included (top spectrum).

**Figure 2.** The normalized pressure slopes, $\Gamma_i = (1/\omega_i)(\partial\omega_i/\partial P)$ of the RBMs in the DWCNTs as a function of their frequency position. Open (solid) circles refer to the primary (secondary) tubes. Red (green) symbols correspond to data recorded using the 676.4 nm (514.5 nm) line of a $Kr^+$ ($Ar^+$) laser, while black circles are data taken from the literature.[15] Triangles denote data corresponding to a SWCNT material produced by the HiPCO process.[12] The thick gray and the black solid lines through the experimental data are guides to the eye, while the blue dotted horizontal line corresponds to the theoretical prediction of the $\Gamma$ parameter for an isolated tube.[12] Inset shows the $\Gamma_i$ values for the internal tubes.

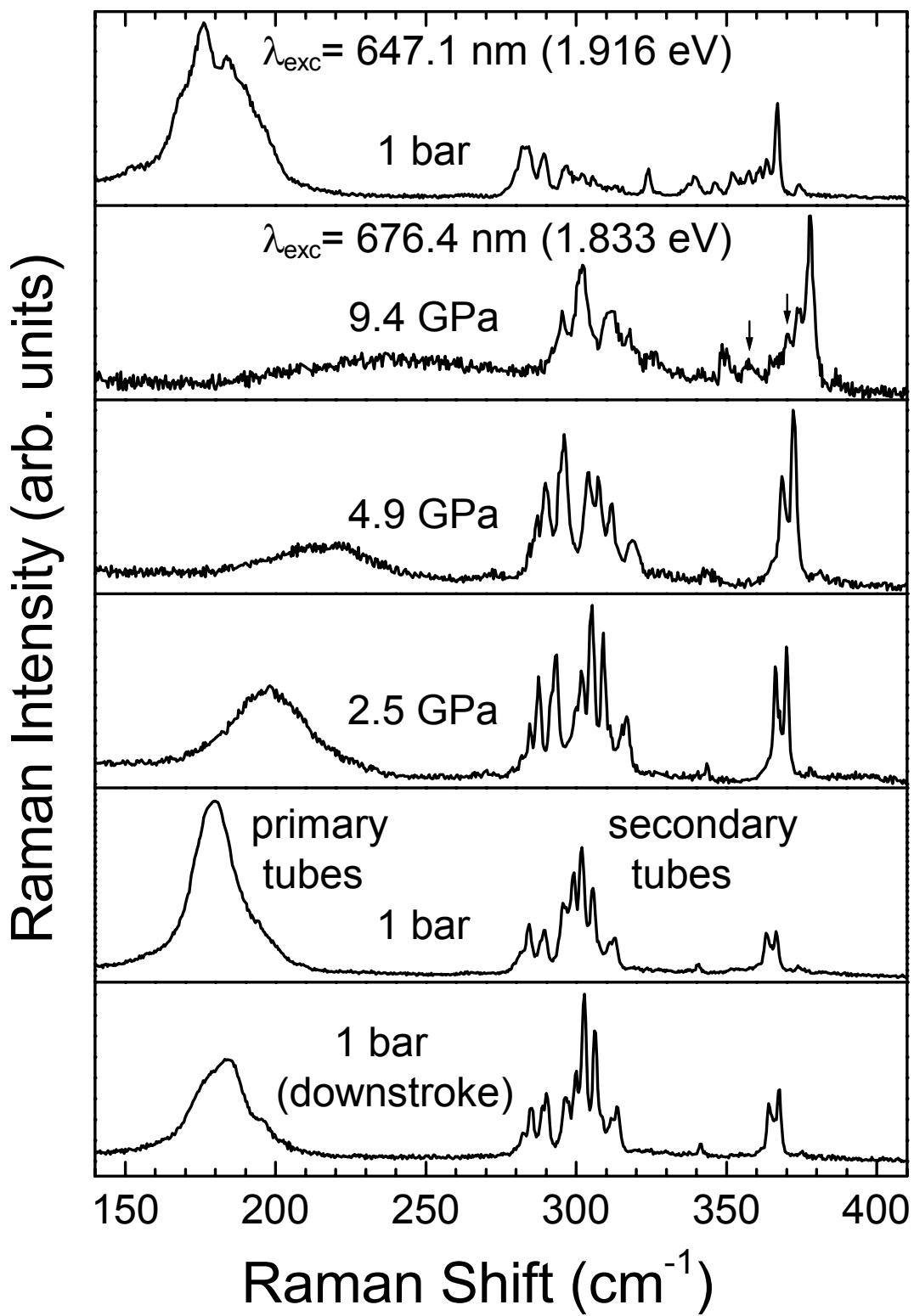

**Fig. 1**

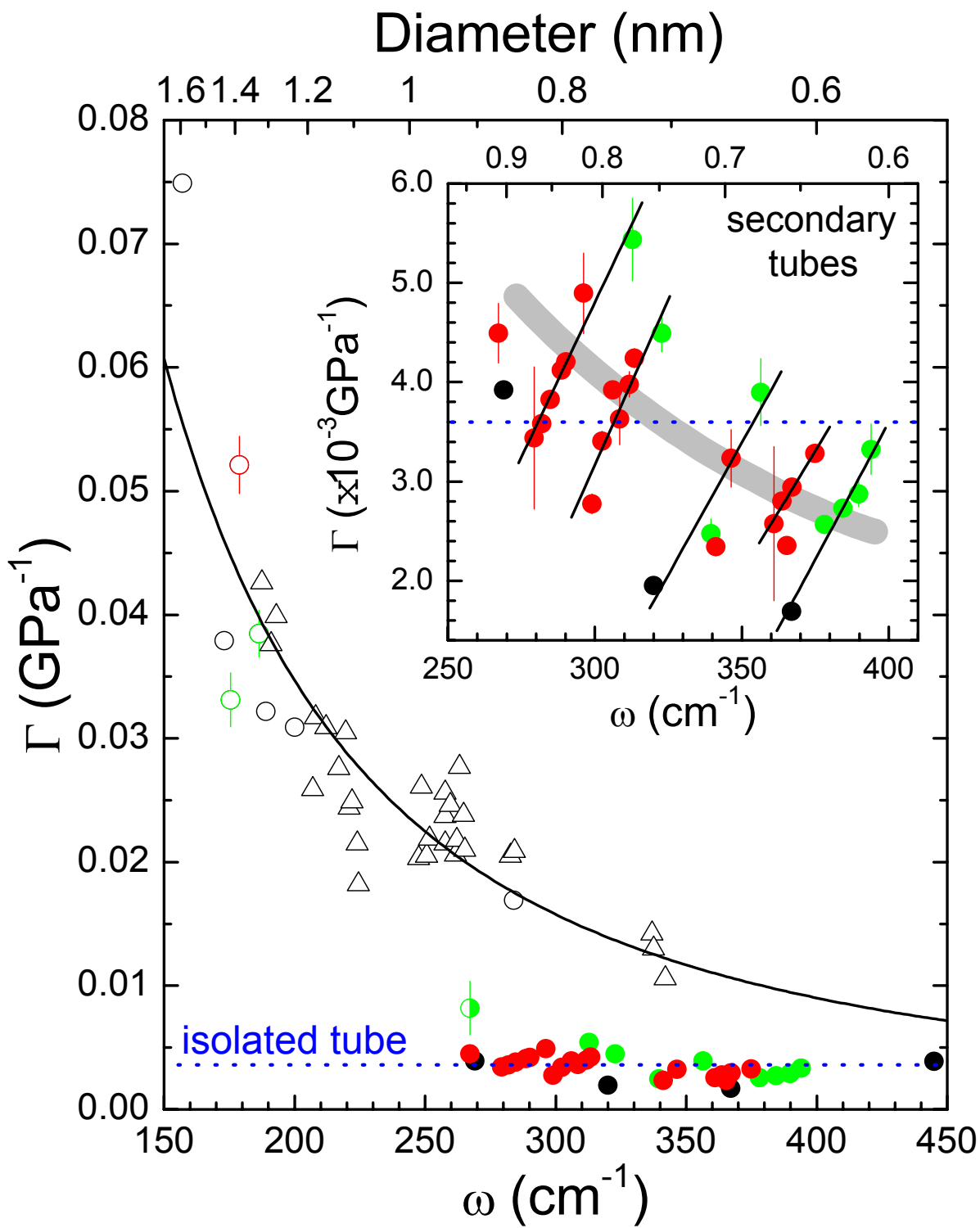

**Fig. 2**